\documentclass[onecolumn,showpacs,preprintnumbers,amsmath,amssymb,aps]{revtex4}
\usepackage{epsfig}

 \normalsize

\def\beq{\begin{equation}}
\def\lsim{\raise0.3ex\hbox{$\;<$\kern-0.75em\raise-1.1ex\hbox{$\sim\;$}}}
\def\gsim{\raise0.3ex\hbox{$\;>$\kern-0.75em\raise-1.1ex\hbox{$\sim\;$}}}
\def\eeq{\end{equation}}
\def\be{\begin{equation}}
\def\ee{\end{equation}}
\def\bea{\begin{eqnarray}}
\def\eea{\end{eqnarray}}

\def\nl{\nonumber \\}
\def\roughly#1{\mathrel{\raise.3ex\hbox
{$#1$\kern-.75em\lower1ex\hbox{$\sim$}}}}

\newcommand{\bers}{\begin{eqnarray*}}
\newcommand{\eers}{\end{eqnarray*}}
\newcommand{\bt}{\begin{itemize}}
\newcommand{\et}{\end{itemize}}

\def\th1{\theta_{LR}}
\def\th2{\theta_{RL}}

\def\ttbar{t \bar{t}}

\def\AFB{A_{FB}}

\begin{document}
\begin{flushright}
UMISS-HEP-2011-03\\
[10mm]
\end{flushright}

\begin{center}
\bigskip {\Large  \bf The Top Forward Backward Asymmetry with general $Z'$ couplings}
\\[8mm]
 Murugeswaran Duraisamy $^{\dag}$ 
\thanks{duraism@phy.olemiss.edu},
Ahmed Rashed $^{\dag, \ddag}$
\thanks{amrashed@phy.olemiss.edu}, and Alakabha Datta $^{\dag}$ 
\thanks{datta@phy.olemiss.edu}
\\[3mm]
\end{center}

\begin{center}
~~~~~~~~~~~ {\it $^{\dag}$ Department  of Physics and Astronomy,}\\ 
~~~~~~~~~~~~{ \it University of Mississippi,}\\
~~~~~~~~~~~~{\it  Lewis Hall, University, Mississippi, 38677, USA}\\
\end{center}

\begin{center}
~~~~~~~~~~~ {\it $^{\ddag}$ Department  of Physics,}\\ 
~~~~~~~~~~~~{ \it Ain Shams University,}\\
~~~~~~~~~~~~{\it  Faculty of Science, Cairo, 11566, Egypt}\\
\end{center}


\date{\today}

\begin{abstract}
The measurement of the top forward-backward asymmetry in $\ttbar$ production measured at the Tevatron shows  deviation from the standard model prediction. A $ u \to t$ transition via a 
flavor-changing $Z^{\prime}$ can explain the data. We show that left-handed $t_Lu_LZ^{\prime}$ couplings  can be  constrained from $B_{d,s}$ mixing  while the constrains on the right-handed couplings  $t_R u_R Z^{\prime}$ vanish in the limit of $m_u \to 0$. We then consider the  most general form of the $t uZ^{\prime}$ interaction which includes  vector-axial vector as well as tensor type couplings and study how these couplings affect the top forward-backward asymmetry.

\end{abstract}
\maketitle
The top quark with its high mass may play a crucial role in electroweak symmetry breaking. Hence the top sector may be sensitive to new physics (NP) effects that could be revealed through careful measurements of top quark properties.
The top quark pair production in proton-antiproton collisions at the Tevatron collider with a  center-of-mass energy of $\sqrt{s} = 1.96$ TeV is dominated by the partonic process $q \bar{q} \to t \bar{t}$. Recently the CDF experiment has  reported a measurement of forward-backward asymmetry in $\ttbar$ production which appears to deviate from the standard model (SM) predictions.
The CDF collaboration  measured the forward-backward asymmetry($A_{FB}$) in top quark pair production in the $t\bar{t}$ rest frame to be $A^{t\bar{t}}_{FB} = 0.475 \pm 0.774$  for 
$M_{t\bar{t}} >450 $ GeV \cite{Aaltonen:2011kc}, which is 3.4 $\sigma$ deviations from the next-to leading order (NLO) SM prediction $A^{t\bar{t}}_{FB} = 0.088 \pm 0.013$
\cite{Kuhn:1998jr,Kuhn:1998kw,Bowen:2005ap,Almeida:2008ug}.
 The D\O\ collaboration also observed a larger than predicted asymmetry~\cite{Abazov:2007qb}. 

\begin{figure}[htb!]
\centering
\includegraphics[width=4.5cm]{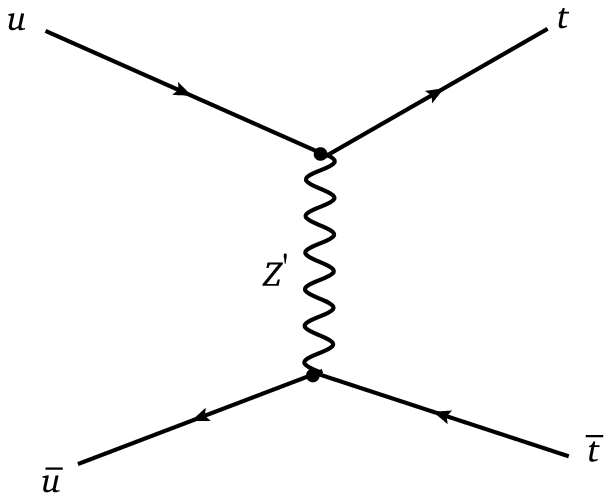}\quad \quad \quad \includegraphics[width=6.0cm]{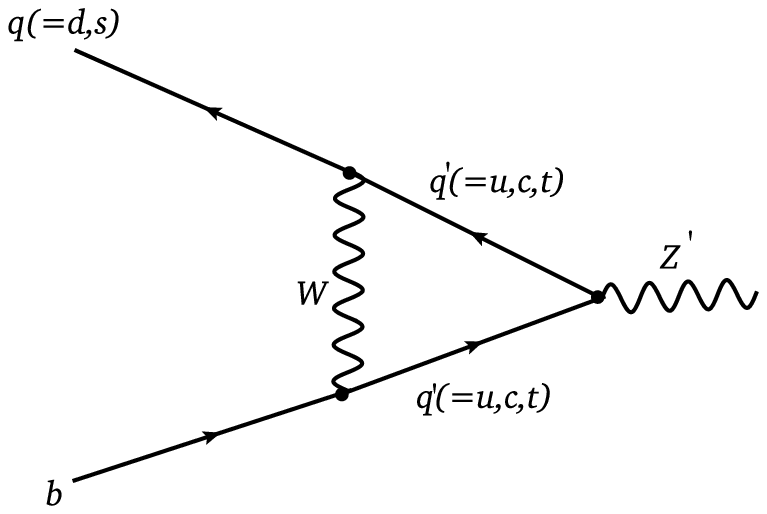}
\caption{ Left panel(a): Tree-level $t\bar{t}$ production diagram involving the $Z^\prime$ exchange. Right panel(b): Tree-level diagram with $tq^{\prime} Z^{\prime}$ coupling ($q^{\prime}=u,c,t$) which generates an effective
$bq Z^\prime$ ( q =d,s) coupling through a vertex correction involving the W exchange.}
 \label{fig.feyndia}
\end{figure}

The current measurement   of the   top quark pair production  cross section from 
4.6 $\mathrm{fb}^{-1}$ of data at CDF is
\bea
\label{CDF-cross}
\sigma_{t\bar{t}} &=& (7.50 \pm 0.48) pb\,,
\eea
for $m_t=172.5$ GeV \cite{Aaltonen:note}, in good agreement with their SM predictions by Langenfeld \emph{et al.} $\sigma_{t\bar{t}}=7.46^{+0.66}_{-0.80}$ pb \cite{Langenfeld:2009wd}, Cacciari \emph{et al.} $\sigma_{t\bar{t}}=7.26^{+0.78}_{-0.86}$ pb \cite{Cacciari:2008zb}, Kidonakis $\sigma_{t\bar{t}}=7.29^{+0.79}_{-0.85}$ pb \cite{Kidonakis:2008mu}, and  recent Ahrens \emph{et al.}'s  significantly low value $\sigma_{t\bar{t}}=6.30 \pm {0.19}^{0.31}_{-0.23}$ pb \cite{Ahrens:2010zv}.
 Hence new physics  models that aim to explain the $\AFB$ measurement must not change the production cross section appreciably. Many NP models that affect $\AFB$, either via s-channel \cite{Sehgal:1987wi,Bagger:1987fz,Ferrario:2009bz,Frampton:2009rk,Jung:2009pi,Chivukula:2010fk,Djouadi:2009nb,Bauer:2010iq,Alvarez:2010js,Chen:2010hm, Delaunay:2011vv, Bai:2011ed, Zerwekh:2011wf,Barreto:2011au,Djouadi:2011aj,Barcelo:2011fw,Westhoff:2011ir,Haisch:2011up,Gabrielli:2011jf} or  $t$-channel exchange of new particles \cite{Jung:2009jz, Cheung:2009ch,Shu:2009xf,Dorsner:2009mq,Arhrib:2009hu,Barger:2010mw, Gupta:2010wt, Xiao:2010hm, Gupta:2010wx,Cheung:2011qa,Cao:2011ew,Berger:2011ua,Barger:2011ih,Grinstein:2011yv, Patel:2011eh,Jung:2011zv,Buckley:2011vc,Shu:2011au,Rajaraman:2011rw,AguilarSaavedra:2011zy,Degrande:2011rt,Wang:2011ta,
 Nelson:2011us,Jung:2011ua,Jung:2011ue,Babu:2011yw,AguilarSaavedra:2011hz} 
have been proposed to explain the forward-backward anomaly. Here we will focus on the model with a $Z^{\prime}$ boson that has a flavor-changing $ t u Z^\prime$ coupling. This coupling can contribute to $\ttbar$ production at the Tevatron via the t-channel exchange of the $Z^{\prime}$ boson (see Fig.~\ref{fig.feyndia}(a)). The $\AFB$ measurement can be explained with a light $Z^{\prime}$ with a mass around 150 GeV and flavor-changing $tuZ^{\prime}$ coupling of $g_{utZ^{\prime}} \sim O(g)$ where $g$ is the weak coupling. One can take higher $Z'$ masses which  requires larger $g_{utZ^{\prime}} \ge 1$ values \cite{zurek}.

Flavor-changing neutral current (FCNC) effects in the SM are tiny and satisfy the condition of natural flavor conservation proposed by Glashow, Weinberg and Paschos \cite{SLGlashow}. The condition of natural flavor conservation can be avoided if quarks of the same charge couple to more than one Higgs or their couplings to a new vector boson  (e.g. a $Z^\prime$ boson) are different for different generation. To date there is no experimental evidence of  FCNC effects beyond those expected from the SM. There are some anomalies in the $B$ system which might require new physics to resolve, but the NP-generated  FCNC effects that  are needed in the $B$ system are much smaller than  the one needed  to resolve the top $\AFB$~\cite{datta}. A tree-level $dbZ^{\prime}$ or a $sbZ^{\prime}$ coupling is strongly suppressed by $B_{d,s}$ mixing. A tree-level $tq^{\prime} Z^{\prime}$ coupling, where $q^{\prime}=u,c,t$, will  generate an effective
$bq Z^\prime( q =d,s)$ coupling through a vertex correction involving the W exchange \cite{zhang} (see Fig.~\ref{fig.feyndia}(b)). The $B_{q}$ mixing constraints on these effective vertices would then lead to constraints on the $tq^{\prime}Z^{\prime}$ coupling. The vertex corrections are divergent and can be regulated by a cut off $\Lambda$, which represents the scale of NP in an effective theory framework.  In NP models where there are no bare $bq Z^\prime$ couplings, the vertex corrections with a chosen $\Lambda$ can be used to constrain the $tq^{\prime}Z^{\prime}$ coupling from $B_{q}$ mixing measurements. We will take the scale of new physics to be $\sim TeV$. In specific complete models $\Lambda$ will represent the mass of some new particles. In models of NP where there are bare $bqZ^{\prime}$ couplings  the vertex correction will renormalize the bare $bqZ^{\prime}$ vertices to produce the renormalized vertices $U_{qb}$. These renormalized vertices can then be fitted to $B_{q}$ mixing data. Assuming the vertex
 corrections to be less than or at most the same size of the bare couplings one, we can obtain   bounds on the $tqZ'$ couplings by requiring the generated $bqZ^{\prime}$ coupling to be $\le U_{qb}$.  It is possible to have models where large bare $bqZ^{\prime}$ couplings cancel with large vertex corrections to produce small renormalized $bqZ^{\prime}$ vertices consistent with experiments. We will not consider these finely tuned model. 
 
When the vertex corrections are computed, one finds that right-handed $tu Z^{\prime}$ couplings do not contribute to $B_q$ mixing in the limit of setting the up quark mass to zero. We note that $ttZ^{\prime}$ couplings do not have such suppression and will contribute to $B_{q}$ mixing via the vertex corrections. Even though the $ttZ^{\prime}$ coupling does not contribute to the top $\AFB$, in specific models of NP this coupling may be related to the FCNC coupling $t u Z^{\prime}$ \cite{Fox}. It turns out the $B_{q}$ mixing constraints on $ttZ^{\prime}$ are  weak because of the small Cabibbo-Kobayashi-Maskawa (CKM) matrix elements $V_{ts(d)}$ and not because of right-handed couplings. The $tq Z^{\prime}(q=u,c,t)$ couplings via box diagrams can produce an effective
$\bar{d} (\bar{s})b \bar{u} u$ operator that can contribute to decays like $ B \to K(K^*) \pi(\eta, \eta'\rho)$ or $B \to \pi(\rho) \pi( \rho)$, etc. decays. The effects of these new operators can be observed in CP-violating and/or triple product measurements \cite{dattaTP}. However, these effective operators only modify the SM Wilson's coefficients in the SM effective Hamiltonian and  so the CP-violating predictions and/or triple product measurements  should be similar to the SM for a reasonable choice of $tq Z^{\prime}(q=u,c,t)$ couplings. 

 We will next consider the most general $tuZ^{\prime}$ couplings including both vector, axial vector and tensor couplings( $ \sim \frac{\sigma_{\mu\nu}q^\nu}{m_t}$) and study the effect of these couplings on the top $\AFB$. 
The interesting feature about these tensor couplings are that we can avoid the $B_q$ mixing constraints due to the suppressions of these operators at low energies~\cite{dattazhang}. The momentum dependence of these operators imply that at the b quark scale these operators will be suppressed by $\sim m_b/m_t$ and consequently the $B_q$ mixing constraints will be weak for these operators.
 
The paper is organized in the following manner. In the next section we  discuss the $B_{q}(q=d,s)$ constraints on the $t u Z^\prime$ operators. In the following section we  introduce the general $tu Z^\prime$coupling including tensor terms and study the effects in the top $\AFB$. This is followed by the section on the $t \to u Z^\prime$ branching ratio calculations. In the final section we present our conclusions. 

\section{ Constraints on $tq^\prime (= u,t) Z^\prime$ couplings from $B_{q(=d,s)}$ mixing}
In general,  new physics contributions to the mass difference between neutral $B_q$ meson mass eigenstates ($\Delta M_q$) can be constrained by the $\Delta M_q$ experimental results. In the SM, $B^0_q$-$\bar{B}^0_q$ mixing  occurs at the one-loop level by the flavor-changing weak interaction box diagrams. The mixing amplitude $M^q_{12}$ is related to the mass difference $\Delta M_q$ via $ \Delta M_q = 2 |M^q_{12}|$. The recent theoretical estimations for the mass differences of $B^0_s$-$\bar{B}^0_s$ and $B^0_d$-$\bar{B}^0_d$ mixing  \cite{Lenz:2010gu} at $1 \sigma$ confidence level are
\bea
\label{eq:delMSM}
(\Delta M_s)^{SM} &=& 16.8 ^{+2.6}_{-1.5}\, \mathrm{ps}^{-1}\,,\quad \quad (\Delta M_d)^{SM} = 0.555 ^{+0.073}_{-0.046} \, \mathrm{ps}^{-1}.
\eea
 The latest measurements of mass difference by CDF \cite{Abulencia:2006ze} and  D\O\ \cite{Abazov:2006dm} for $B_s$ mixing are
\bea %
\label{MBSexp}
\Delta M_{B_s} &=& (17.77\pm 0.10 ({\rm stat.}) \pm 0.07({\rm syst.}) )\,\mathrm{ps^{-1}}\, \nl
\Delta M_{B_s} &=& (18.53\pm 0.93 ({\rm stat.}) \pm 0.30({\rm
syst.})) \,\mathrm{ps^{-1}}\,.
\eea %
The Heavy Flavor Averaging Group value for the mass difference of  $B^0_d$-$\bar{B}^0_d$ mixing is   $\Delta M_{B_d} \mathrm{(exp)} = (0.507 \pm 0.004)\,\mathrm{ps^{-1}}$ \cite{Asner:2010q}. The experimental results for the mass differences of both $B^0_s$-$\bar{B}^0_s$ and $B^0_d$-$\bar{B}^0_d$ mixing  are  consistent with their SM expectations. Hence, the mass difference results can provide  strong constraints on NP contributions.

In this section we will consider the $B_{d,s}$  mixing constraints on the $tq^\prime (= u,t) Z^\prime$ couplings. 

\subsubsection{$tu Z^\prime$  left-handed coupling}
The most general Lagrangian for  flavor-changing $tuZ^\prime$ transition is \cite{dattamurgesh}
\bea
\label{Lagtuzp-def}
{\cal{L}}_{tu Z^\prime} &=&  \bar{u} \Big[ \gamma^\mu (a+ b\gamma_5)  + i  \frac{\sigma_{\mu \nu}}{m_t}  q^\nu (c+d \gamma_5) \Big]t Z^{\prime}_\mu \,, 
\eea

where $q = p_t - p_u$. In general, the couplings a, b, c and d are complex and can be momentum-dependent (form factors). In this work we will take the couplings to be constants with no momentum dependence. Consider the $tuZ^\prime$ vertex with  $a= -b = g^L_{tu}$ , and c=d=0 in Eq.~(\ref{Lagtuzp-def}). This  generates effective  $bq Z^\prime( q =d,s)$ coupling  at one-loop level due to  W exchange.  We obtain the $bq Z^\prime$ coupling in the Pauli-Villars regularization as
 
\bea
{\cal{L}}_{Z^\prime} = U_{qb} \bar{q} \gamma^\mu (1-\gamma_5) b Z^\prime_\mu \,,
\eea
where 
\bea
\label{U-def}
 U_{qb} & =&  g^L_{tu} \, \frac{G_F}{\sqrt{2}} M^2_W (V^*_{uq} V_{tb} + V^*_{tq} V_{ub}) \frac{1}{8 \pi^2}\Big[\frac{ x_t  Log[\frac{\Lambda^2}{m^2_t}] -Log[\frac{\Lambda^2}{M^2_W}] }{(x_t-1)}\Big]\,.
\eea
where $ \Lambda \sim $ TeV is a cutoff scale, and $x_t = m^2_t / M^2_W$. The function $U_{qb}$ includes only the contribution from the W boson, and the contribution of the associated Goldstone boson in the SM is the order of $m_u/M_W$. Note that for $B_d$ mixing the coupling $g^L_{tu}$ is associated with the CKM factor $ V^*_{ud} V_{tb} \sim 1$, and thus one can expect a strong constraint on  $g^L_{tu}$ from the mass difference $\Delta M_d$.

A tree-level exchange of the  $Z^\prime$ generates  the $\Delta B =2$ effective Lagrangian responsible for the neutral $B_q$ meson mixing
\bea
{\cal{H}}^{\Delta B =2}_{Z^\prime} = \frac{U_{qb}^2}{M^2_{Z^\prime}} \eta_{Z^\prime} (\bar{q} b)_{V-A}(\bar{q} b)_{V-A}\,,
\eea
where $(\bar{q} b)_{V-A}=\bar{q} \gamma^\mu (1-\gamma_5) b $, and the QCD correction factor $\eta_{Z^\prime} = [\alpha_s(M_{Z^\prime})/\alpha_s(m_{b})]^{6/23}$. The  $Z^\prime$ contribution to the $B_q$ mixing amplitude can be obtained by using the vacuum insertion method as
\bea
\label{eq:M12qZp}
[M^{q}_{12}]^{Z^\prime} &=& \frac{4}{3}\frac{U_{qb}^2}{M^2_{Z^\prime}} \eta_{Z^\prime} m_{B_q} f^2_{B_q} B_q.
\eea

In the presence of new physics, the mixing amplitude $M^{q}_{12}$ can be parameterized by complex parameters $\Delta_{q}$ \cite{Lenz:2010gu}  

\bea
\label{Delq-def}
M^{q}_{12} = [M^{q}_{12}]^{SM} \Delta_q.
\eea

In our case, $\Delta_q = |\Delta_q| e^{i \phi^{\Delta}_q} = 1+[M^{q}_{12}]^{Z^\prime}/[M^{q}_{12}]^{SM}$. A global analysis on the parameters $|\Delta_q|$ and $ \phi^{\Delta}_q$ for  $B_d-\bar{B}_d$ and $B_s-\bar{B}_s$ mixing are carried out in \cite{Lenz:2010gu}. The best fit results for $\Delta_d$ and $\Delta_s$ in this analysis  at 1 $\sigma$ confidence level (scenario I) are 
\bea
\label{delad-lenz}
|\Delta_d|= 0.747^{+0.195}_{-0.082} , \quad \quad \phi^{\Delta}_d = -12.9^{+3.8^\circ}_{-2.7},
\eea
and
\bea
\label{delas-lenz}
|\Delta_s|= 0.887^{+0.143}_{-0.064}, \quad \quad \phi^{\Delta}_s =  -51.6^{+14.2^ \circ}_{-9.7} \quad \mathrm{or} \quad -130.0^{+13^\circ}_{-12}.
\eea

\begin{figure}[htb!]
\centering
\includegraphics[width=5.5cm]{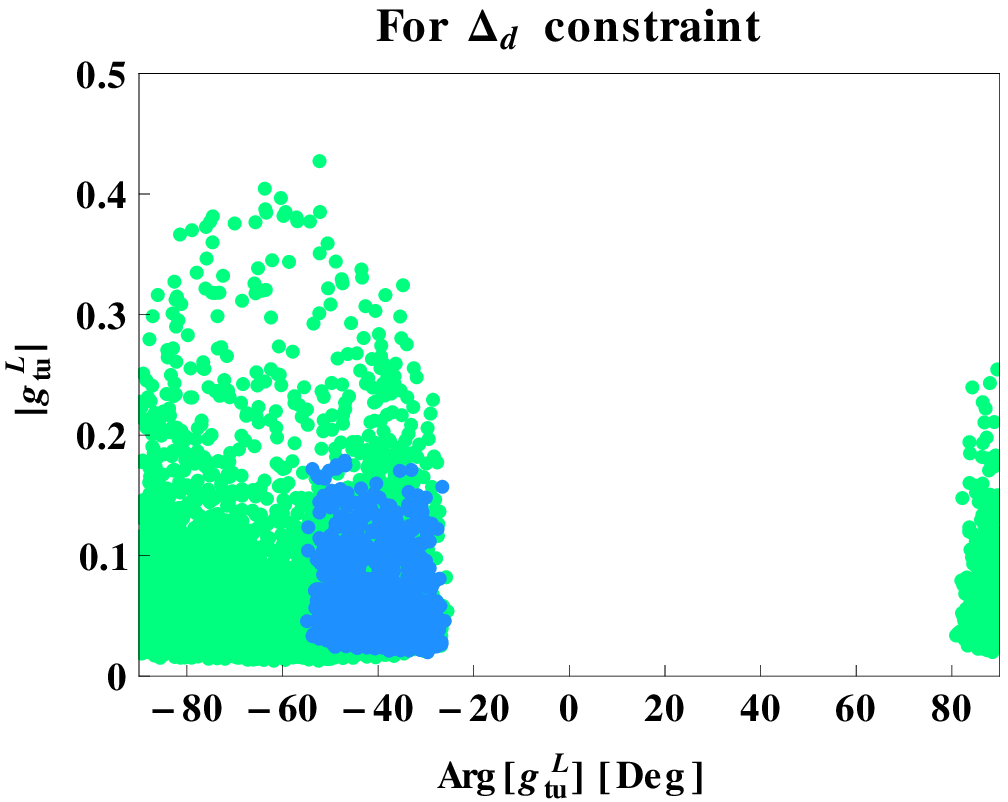}\quad \quad \includegraphics[width=5.5cm]{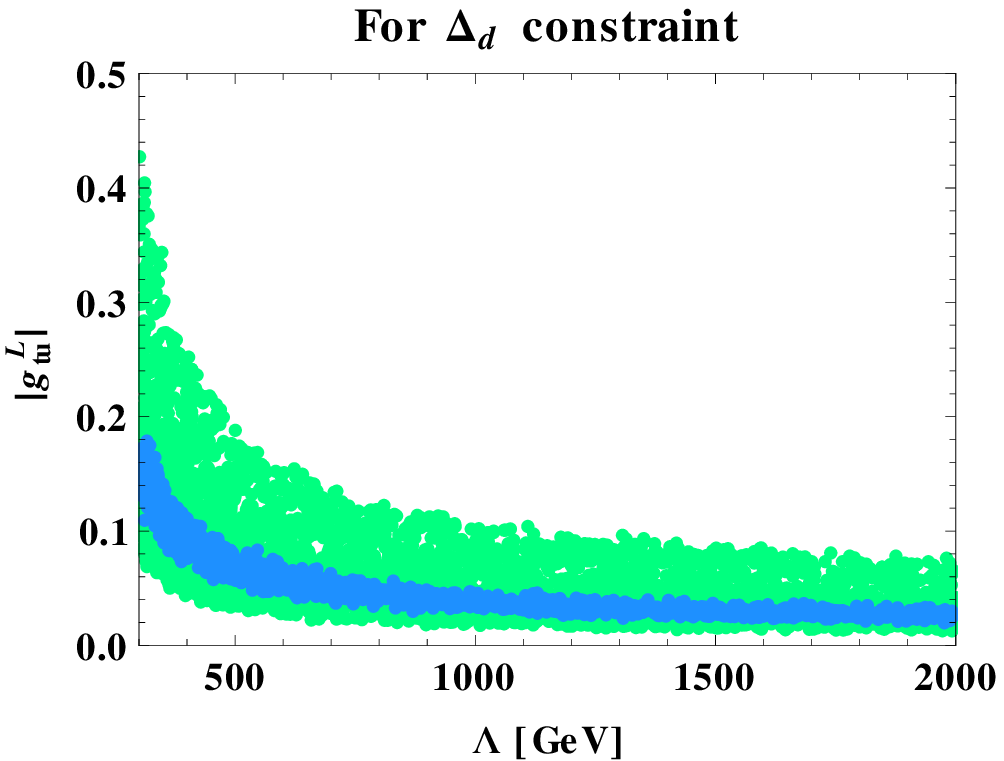}
\caption{ $|g^L_{tu}|$ vs Arg[$g^L_{tu}$][Deg] (left panel) and $|g^L_{tu}|$ vs $\Lambda$[GeV] (right panel) for $B_d$ mixing. Green scatter points are constrained by  $|\Delta_d|$. Blue scatter points are constrained by  $|\Delta_d|$ and $\phi^{\Delta}_d$.}
 \label{fig.Bdconst}
\end{figure}

The $\Delta_d $  constraint in Eq.~(\ref{delad-lenz}) on the coupling $g^L_{tu}$ at $\bar{m_t}(\bar{m_t}) = (165.017 \pm  1.156 \pm 0.11)$ GeV \cite{Lenz:2010gu}, $\beta^{SM} = 27.2^{+1.1^\circ}_{-3.1} $ \cite{Lenz:2010gu}, and $M_{Z^\prime} =150$ GeV is shown in Fig.~\ref{fig.Bdconst}. The numerical values of all other theoretical inputs can be found in \cite{Lenz:2010gu}. They are varied  within 1 $\sigma$ errors in the fit. The cutoff scale $\Lambda $ is varied between 300 GeV to 2 TeV . The green scatter points in Fig.~\ref{fig.Bdconst} satisfy only $|\Delta_d| $ in Eq.~(\ref{delad-lenz}), while blue points satisfy both $|\Delta_d|$ and $\phi^{\Delta}_d$ in Eq.~(\ref{delad-lenz}). The results  
indicates  that $B_d$ mixing can strongly constrain the $tuZ^\prime$ coupling $g^{L}_{tu}$ even at $\Lambda = 300 $ GeV. In particular we note that the maximum value for 
$|g^L_{tu}|$ is around 0.2 and is associated with a large phase. In fact there are no real
$g^L_{tu}$ that satisfy the $B_d$ constraint.

\begin{figure}[htb!]
\centering
\includegraphics[width=5.5cm]{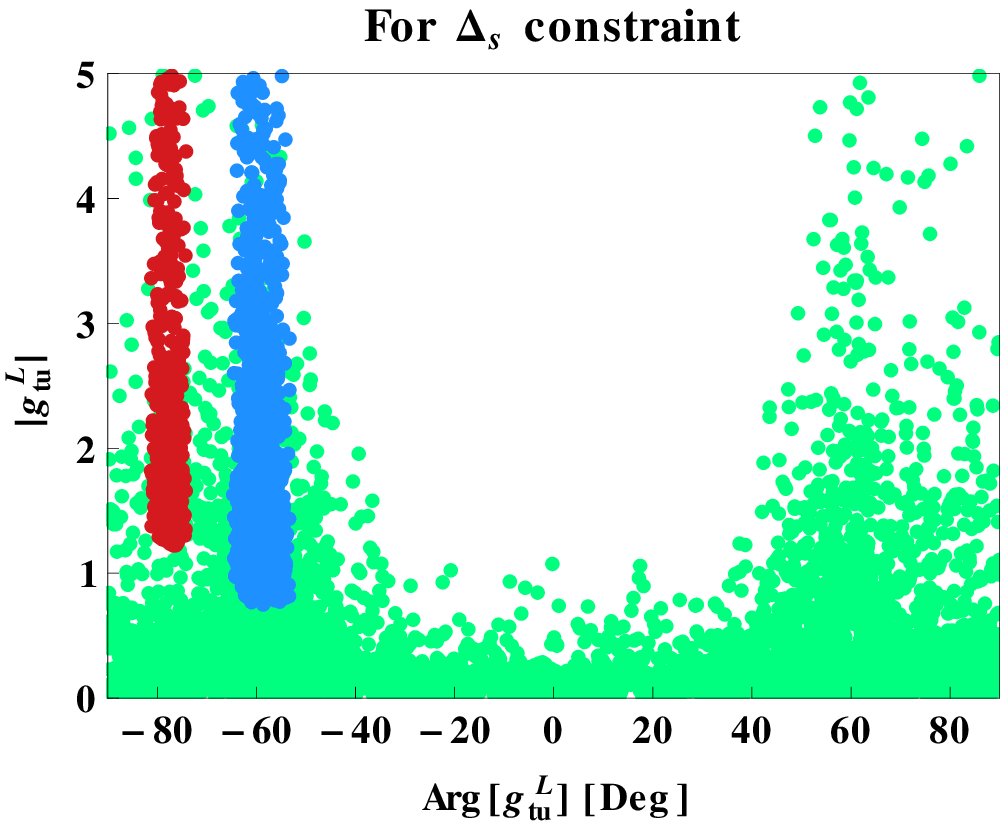}\quad \quad \includegraphics[width=5.5cm]{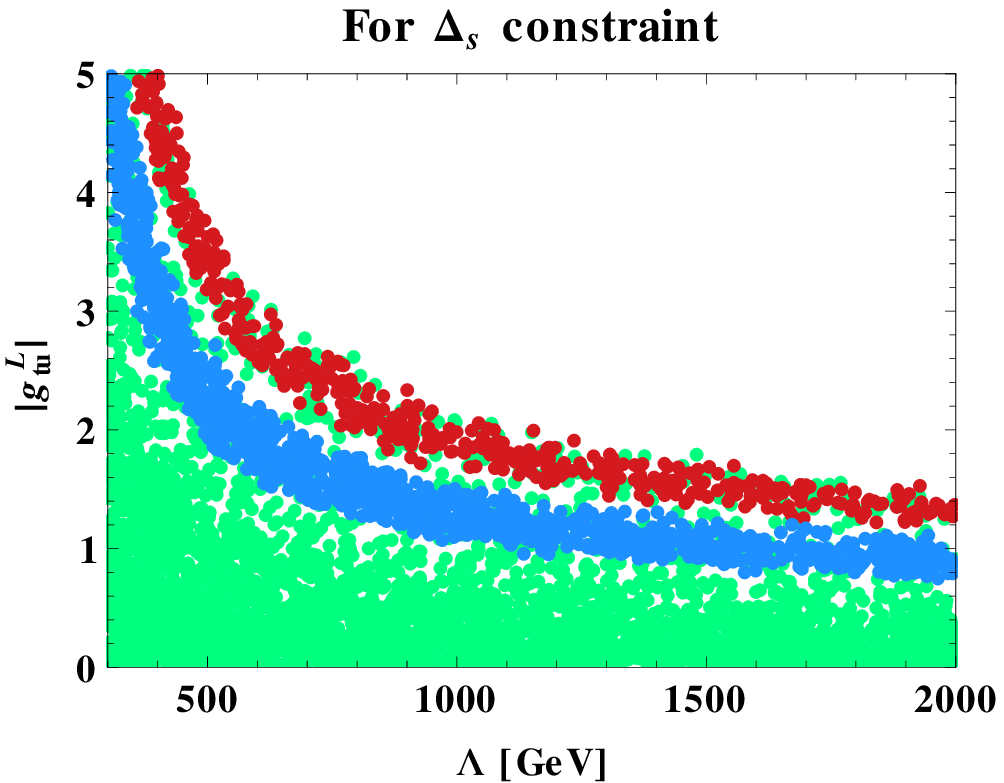}
\caption{ $|g^L_{tu}|$ vs Arg[$g^L_{tu}$][Deg] (left panel) and $|g^L_{tu}|$ vs $\Lambda$[GeV] (right panel) for $B_s$ mixing. Green scatter points are constrained by  $|\Delta_s|$. Blue scatter points are constrained by  $|\Delta_d|$ and $\phi^{\Delta}_d = -51.6^{+14.2^ \circ}_{-9.7}$. Red scatter points are constrained by  $|\Delta_d|$ and $\phi^{\Delta}_d = -130.0^{+13^\circ}_{-12}$.}
 \label{fig.Bsconst}
\end{figure}

On the other hand, Fig.~\ref{fig.Bsconst} suggests that the constraints from $B_s$ mixing  
on the $tuZ^\prime$ coupling $g^{L}_{tu}$ are weaker  ($\sim O(1)$) even  at $\Lambda = 2 $ TeV.
This can be understood from the fact that the $B_s$ mixing contribution in this case 
is associated with the CKM factor $V^*_{us} V_{tb}$ and is suppressed.
 The (green, blue, red) scatter points in Fig.~\ref{fig.Bsconst} are constrained by ( $|\Delta_s| $,  \{$|\Delta_s|$, $\phi^{\Delta}_s = -51.6^{+14.2^ \circ}_{-9.7}$\}, \{$|\Delta_s|$, $\phi^{\Delta}_s = -130.0^{+13^\circ}_{-12}$\}) in Eq.~(\ref{delas-lenz}), respectively. The large negative phase $\phi^{\Delta}_s$ prefers large $g^{L}_{tu}$ values.

\subsubsection{$tu Z^\prime$  right-handed coupling}
We now consider the $tuZ^\prime$ vertex with right-handed couplings,  $a= b= g_R$ , and c=d=0. The contribution of this vertex to  $M_{12}$  is   suppressed by $m^2_u/m^2_W$. Hence, the right-handed coupling $g_R$ cannot be constrained by $B_q$ mixing.

Finally as indicated in the earlier section, the left- and the right-handed couplings generate via the box diagram  effective $\bar{q} b \bar{u}u$( $q=d,s$) operators. These operator can be constrained by observables in nonleptonic B meson decays like $ B \to \pi \pi/ K \pi$. These operators change the Wilson's coefficients of the SM effective Hamiltonian with the change being $\sim 10^{-2}$ at the scale $\mu = M_W$ for $M_{Z^\prime} = 150 $ GeV and $g^L_{tu},g^R_{tu} \sim O(g)$.  Since the generated NP physics operator structures are similar to the SM there are no easy way to detect their presence. A detailed fit to all the nonleptonic data may provide constraints on the
couplings $g^{L,R}_{tu}$, which we do not perform in this work.
Some analysis along this line has been  done for  $tdW^\prime$ coupling in  \cite{Chen:2011mga}.  

\subsubsection{$tt Z^\prime$   coupling}

For completeness, next we consider  $B_q$ mixing  constraints on the $tt Z^\prime$  couplings. The Lagrangian for the $ttZ^\prime$ interaction is 
\bea
\label{eq:LagttZp}
{\cal{L}}_{ttZ^{\prime}} &=&\bar{t}[ g^L_{tt} \gamma^\mu (1-\gamma_5)+ g^R_{tt}  \gamma^\mu(1+\gamma_5) ]t Z^{\prime}_\mu.
\eea

Again, we evaluate the one-loop diagram (see  Fig.~\ref{fig.feyndia}(b))  in the Pauli-Villars regularization and obtain the effective Lagrangian for bq(=d,s)$Z^\prime$ interaction as
 
 \bea
\label{eq:LagUUZ}
{\cal{L}}^\prime_{Z^{\prime}} &=& U^\prime_{qb} \bar{q} \gamma^\mu (1-\gamma_5) b  Z^{\prime}_\mu \,, 
\eea 
where
\bea
\label{eq:ugbtt}
U^\prime_{qb} &=&  \frac{G_F}{\sqrt{2}} M^2_W V_{t q} V_{tb} f_{tt}(\Lambda,x_t)\,,
\eea
with
\bea
\label{eq:ftt}
 f_{tt}(\Lambda,x_t) &=& \frac{1}{(4 \pi^2)} \int^1_0 dx \int^{1-y}_0 dy \Big[g^{L}_{tt} \Big(Log[\frac{x \Lambda^2}{M^2_W D_{tt}}]+\frac{1}{2} \frac{x^2_t}{ D_{tt}}\Big)+ g^{R}_{tt}  x_t \Big(\frac{1}{2} Log[\frac{x \Lambda^2}{M^2_W D_{tt}}]+\frac{1 }{ D_{tt}}\Big) \Big]\,,
\eea
and $D_{tt}= x + (1-x) x_t$. The function $f_{tt}$ includes both the W boson and the associated Goldstone boson contributions. The  $ttZ^\prime$ contribution to the $B_q$ mixing amplitude is
\bea
\label{eq:M12qttZp}
[M^{q}_{12}]^{Z^\prime} &=& \frac{4}{3}\frac{[U^\prime_{qb}]^2}{M^2_{Z^\prime}} \eta_{Z^\prime} m_{B_q} f^2_{B_q} B_q~.
\eea

 Both  $B_d-\bar{B}_d$ and $B_s-\bar{B}_s$ constraints in Eqs.~(\ref{delad-lenz}) and (\ref{delas-lenz}) can allow large $ \sim O(1)$ values for $g^{L,R}_{tt}$.

\section{Top quark forward-backward asymmetry}
In this section we calculate the top $\AFB$ keeping in mind the constraints derived on the coupling from the previous section.
The most general Lagrangian for a flavor-changing $tuZ^\prime$ interaction is given in Eq.~(\ref{Lagtuzp-def}). This interaction can contribute to $u \bar{u} \to t \bar{t}$ scattering amplitude through the t-channel exchange of the $Z^\prime$ boson (see Fig.~\ref{fig.feyndia}(a)).  The tree-level differential cross section for $ q \bar{q} \to t \bar{t}$  process in the $t\bar{t}$ center-of-mass frame including both the SM and for $Z^\prime$ contributions is
\bea
\label{dc-part} 
\frac{d \hat{\sigma}}{d \cos{\theta}} &=& \frac{\beta_t}{ 32 \pi \hat{s}} ( {\cal{A}}_{SM}+ {\cal{A}}_{SM-Z^\prime}+ {\cal{A}}_{Z^\prime})\,,
\eea

where  $\hat{s}= (p_{q}+ p_{\bar{q}})^2$ is the squared center-of-mass energy of the $ t \bar{t}$ system, $\beta_t= \sqrt{1-4 m^2_t/\hat{s}}$, and the polar angle $\theta$  is the relative angle between direction  of motion of the outgoing top quark and  the incoming q quark. The quantities  ${\cal{A}}_{SM}$, ${\cal{A}}_{SM-Z^\prime}$, and $ {\cal{A}}_{Z^\prime}$ denote the leading order SM, the interference between the SM and $Z^\prime$, and the pure $Z^\prime$ scattering amplitudes, respectively. These amplitudes can be obtained in terms of kinematic variables $\theta$ and $\hat{s}$ as

\bea
\label{Sc-amp}
{\cal{A}}_{SM} &=& \frac{2 g^4_s}{9} \Big[1+ c^2_\theta + \frac{4 m^2_t}{\hat{s}} \Big]\,, \nl
{\cal{A}}_{SM-Z^\prime}  &=& \frac{2 g^2_s}{9} \Big[\frac{\hat{t}-M^2_{Z^\prime}}{(\hat{t}-M^2_{Z^\prime})^2+ M^2_{Z^\prime} \Gamma^2_{Z^\prime}}\Big] (f_1 + f_2)\,, \nl
{\cal{A}}_{Z^\prime}  &=& \frac{1}{4} \Big[\frac{1}{(\hat{t}-M^2_{Z^\prime})^2+ M^2_{Z^\prime} \Gamma^2_{Z^\prime}}\Big](f_3+f_4+f_5).
\eea
Where  $c_\theta = \beta_t \cos{\theta}$, and $\hat{t}= (p_q -p_t)^2 = -\hat{s}/2 (1-\beta_t \cos{\theta}) + m^2_t$. The functions $f_i$s (i = 1-5)   can be found in the Appendix. Here we assume the couplings a, b, c and d to be real. Our results for $\bar{t} t$ production are obtained by the convolution of the analytic differential cross section of Eq.~(\ref{dc-part})  with the  CTEQ-5L parton distribution functions \cite{Lai:1999wy} implemented in Mathematica. We expect the MSTW 2008 \cite{Martin:2009iq} parton distributions to give compatible results.

The forward-backward asymmetry of the top quark in the  $t\bar{t}$ c.m. frame is defined as \cite {Cao:2010zb}

\bea
\label{AFT-def}
A^{t\bar{t}}_{FB} &=& \frac{\sigma_F -\sigma_B}{\sigma_F +\sigma_B}\,,
\eea
where  
\bea
\label{sigfB-def}
\sigma_F &=& \int^1_0 \frac{d \sigma}{d \cos{\theta}} d \cos{\theta}\,, \quad \sigma_B = \int^0_{-1} \frac{d \sigma}{d \cos{\theta}} d \cos{\theta}.
\eea

In our analysis, we choose some representative values for the couplings a, b, c, and d to generate large forward-backward asymmetry $A^{t\bar{t}}_{FB}$   for high $M_{t\bar{t}} $ ($> 450$ GeV) without distorting the shape of the mass spectrum $d\sigma_{t\bar{t}}/dM_{t\bar{t}}$. We fix the renormalization and factorization scales at $\mu_R = \mu_F = m_t$. We evaluate  $A^{t\bar{t}}_{FB}$ which includes the  NLO SM  and the $Z^\prime$ contributions at $m_t =172.5$ GeV. Also, we apply a QCD K-factor K = 1.3 to the tree-level cross section  in order to match the  SM prediction for $\sigma_{t\bar{t}} $. We consider the  $Z^\prime$ boson with mass $M_{Z^\prime}$ = 150 GeV and width $\Gamma_{Z^\prime} =0$ for the numerical analysis.

\subsection{Pure vector-axial vector couplings: $a= \mp b$ and  $c= d=0$}

\begin{figure}[htb!]
\centering
\includegraphics[width=5.5cm]{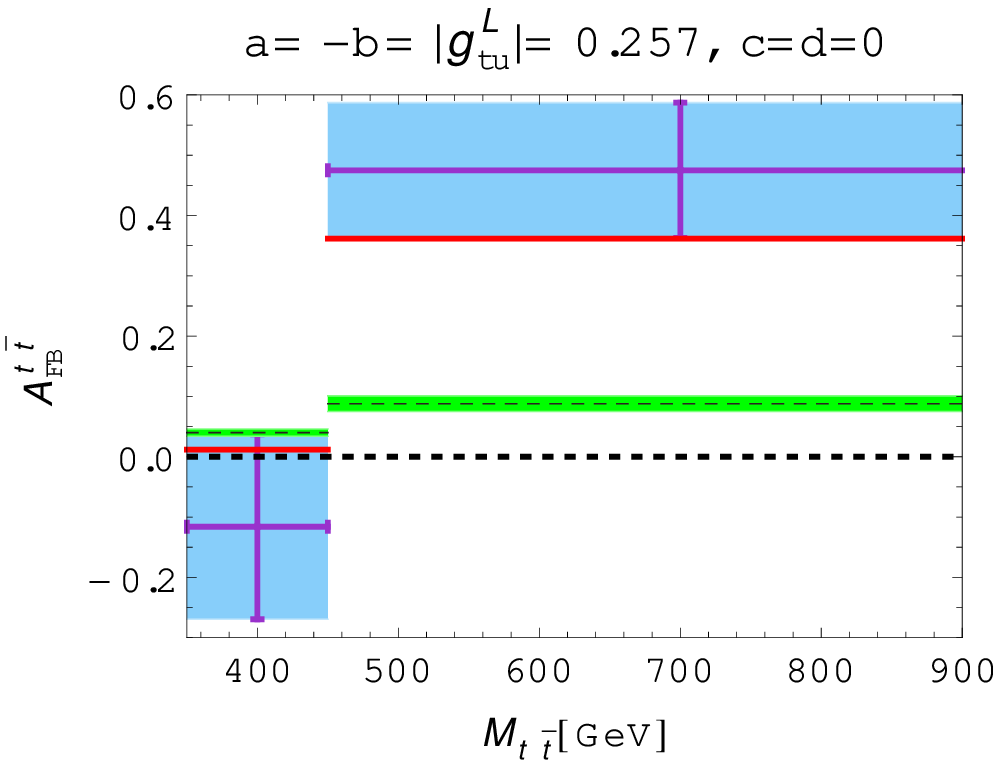}\quad \quad \includegraphics[width=5.5cm]{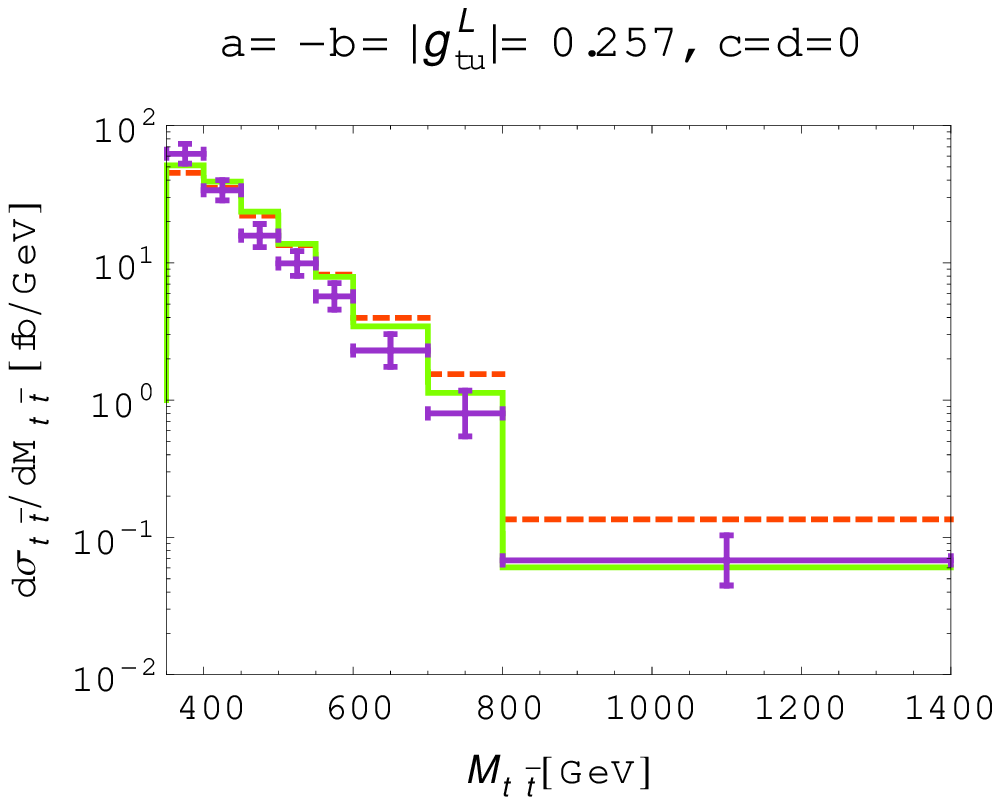} 
\caption{Left panel: $M_{t\bar{t}}$ distribution of $A^{t\bar{t}}_{FB}$ in the two energy ranges [350,450]GeV and [450,900]GeV of invariant mass $M_{t\bar{t}}$. Green band: the SM prediction. Blue band with $1 \sigma$ error bars: the unfolded CDF measurement \cite{Aaltonen:2011kc}. Red line: the SM  with  $Z^\prime$ exchange  prediction for  $( a = -b =  0.257, c = d = 0)$. Right panel: $M_{t\bar{t}}$ distribution of $d\sigma_{t\bar{t}}/dM_{t\bar{t}}$ [in fb/GeV] for eight different energy bins of $M_{t\bar{t}}$. Green line: the NLO SM prediction. Blue band with $1 \sigma$ error bars: the unfolded CDF measurement \cite{Aaltonen:2009iz}. Red line: the SM  with  $Z^\prime$ exchange  prediction for above values of couplings at $m_t = 175$ GeV. }
 \label{fig.AFBonlya}
\end{figure}

This case has already been considered before \cite{Jung:2009jz}, but  only right-handed couplings were considered. Here we will consider both right- and left-handed couplings. 
 We take the representative values of the couplings $a = -b = |g^L_{tu}|= 0.257$, and $c = d =0$.
  This value for $g^L_{tu}$ satisfies the $|\Delta_d|$ constraint but not the phase $\phi^{\Delta}_d$ constraints from $B_d$ mixing (see Fig.~\ref{fig.Bdconst}). For these values $A^{t\bar{t}}_{FB}$ can be explained within one $\sigma$ error of its measurement for $M_{t\bar{t}} > 450 $ GeV.
In Fig.~\ref{fig.AFBonlya}, we show the $M_{t\bar{t}} $ distribution for the $t\bar{t}$ observables $A^{t\bar{t}}_{FB}$, and $\sigma_{t\bar{t}}$. The differential distribution, $d\sigma_{t\bar{t}}/dM_{t\bar{t}}$, has been measured in eight different energy bins of $M_{t\bar{t}}$ for $m_t = 175$ GeV in Ref.~\cite{Aaltonen:2009iz}. Our distribution of $d\sigma_{t\bar{t}}/dM_{t\bar{t}}$   is  consistent with the measurements. Since the partonic scattering amplitudes in this case (see the Appendix) depends  on $b^2$ and $b^4$ terms, our results hold for right-handed couplings also, i.e $a = b = |g^R_{tu}|= 0.257$, and $c = d =0$. 

\subsection{General case: all couplings are present}
In this section  we consider the most general $tuZ^\prime$ couplings. 
 We showed earlier that the  left-handed coupling are strongly constrained  from $B_d$ mixing and  there are no real values of $g^L_{tu}$ that satisfy the $B_d$ mixing constraint.
We now investigate the effect of the couplings $c$ and $d$ on the $\AFB$ predictions. 

\subsection{Pure tensor couplings : $ a = b = 0$ and $c = \pm d$}
\begin{figure}[htb!]
\centering
\includegraphics[width=5.5cm]{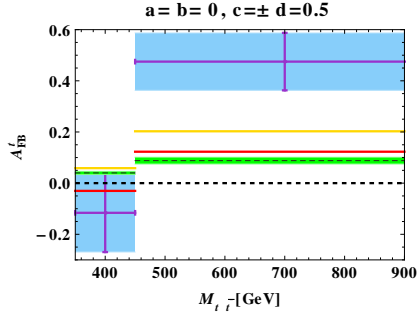} 
\caption{ $M_{t\bar{t}}$ distribution of $A^{t\bar{t}}_{FB}$. Green band: The SM prediction. Blue band with $1 \sigma$ error bars : CDF measurement. Red  and yellow lines: The SM  with  $Z^\prime$ exchange  prediction at $M_{Z^\prime} = 150 $ GeV, and  $M_{Z^\prime} = 100 $ GeV, respectively, for $a=b=0$ and $c = \pm d =0.5$. }
 \label{fig.AFBonlyc}
\end{figure}

We consider the case of pure tensor couplings. In this scenario we can avoid the $B_q$ mixing constraints as the effects of the tensor couplings are suppressed by ${m_b \over m_t}$ at the $b$ mass scale. 
The SM and $Z^\prime$ interference contribution ${\cal{A}}_{SM-Z^\prime}$ in Eq.~(\ref{Sc-amp}) vanishes in this case. The functions $f_4$ and $f_5$ in pure $Z^\prime$ contribution ${\cal{A}}_{Z^\prime}$ are  also zero, and $f_3
$ is order of $(c\hat{s}/m_t)^2$. The mass spectrum for $A^{t\bar{t}}_{FB}$ is shown in Fig. \ref{fig.AFBonlyc}(a) for only $ c = \pm d$ couplings ($c = \pm d = 0.5$). The results indicate that $Z^\prime$ contribution cannot reproduce the $\AFB$ measurement within one $\sigma$ for $M_{t\bar{t}} > 450$ GeV  even at a low $M_{Z^\prime} = 100 $ GeV (yellow lines) value.

\subsection{All the couplings are same order}
\begin{figure}[htb!]
\centering
\includegraphics[width=5.5cm]{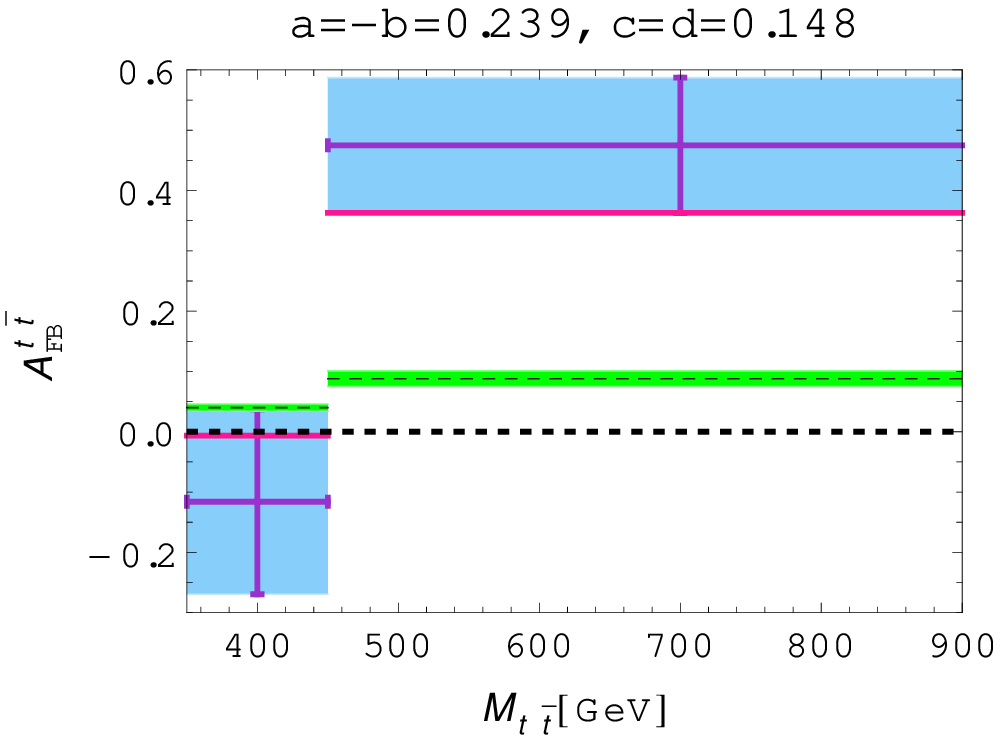}\quad \quad \includegraphics[width=5.5cm]{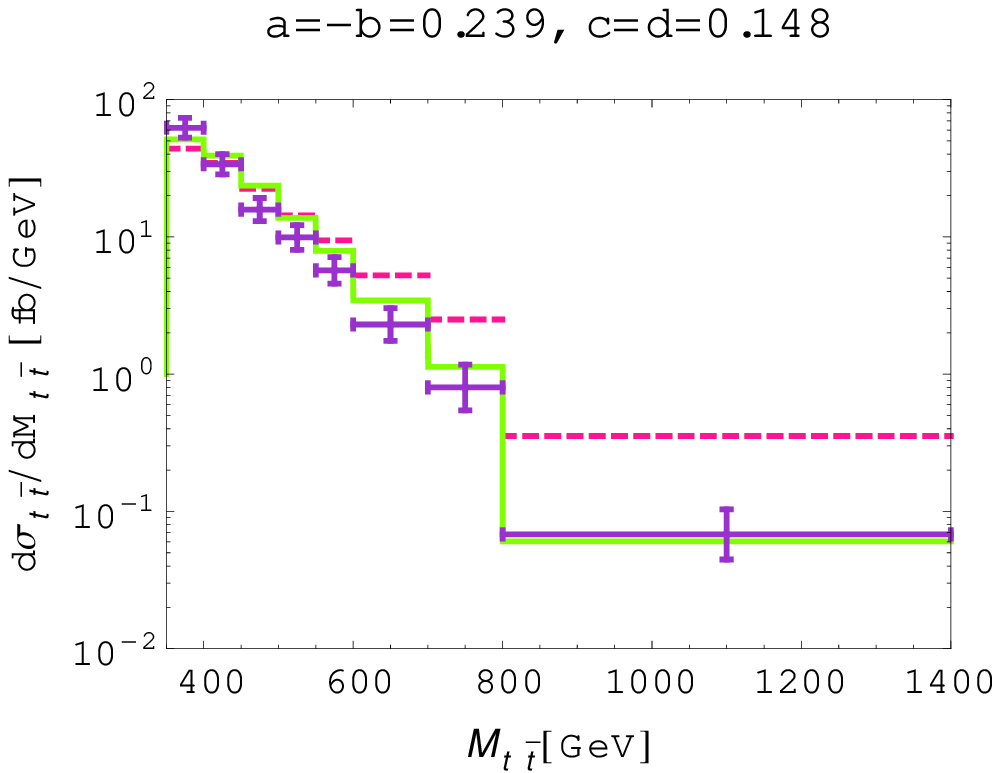} 
\caption{ $M_{t\bar{t}}$ distributions of $A^{t\bar{t}}_{FB}$ and $d\sigma_{t\bar{t}}/dM_{t\bar{t}}$ [in fb/GeV]. Pink lines: the SM  with  $Z^\prime$ exchange  prediction for  ($ a =-b= 0.239$, $c=d = 0.148$). The same conventions as in  Fig.~\ref{fig.AFBonlya}  used for other lines. }
 \label{fig.AFBaeqmbceqd}
\end{figure}

\begin{figure}[htb!]
\centering
\includegraphics[width=5.5cm]{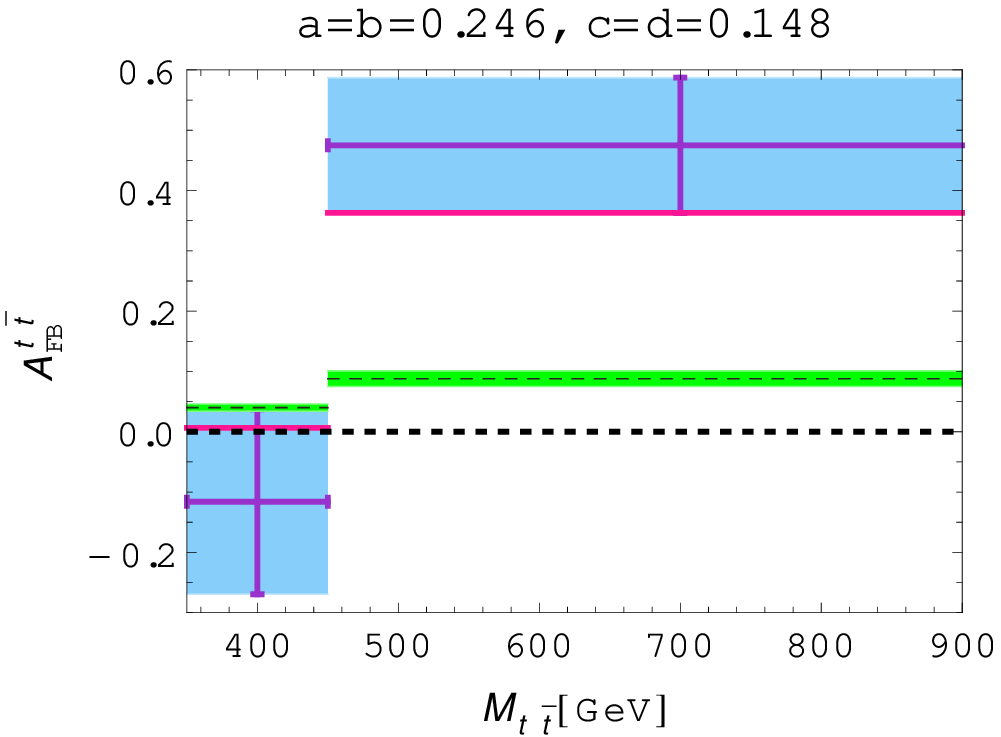}\quad \quad \includegraphics[width=5.5cm]{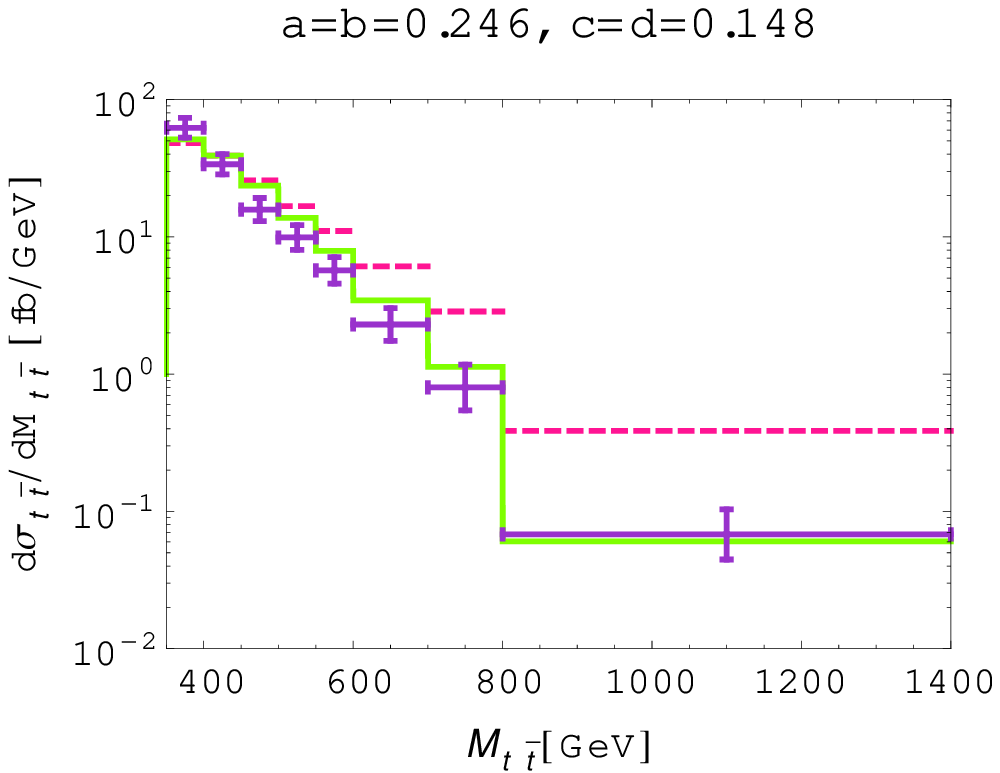} 
\caption{ $M_{t\bar{t}}$ distributions of $A^{t\bar{t}}_{FB}$ and $d\sigma_{t\bar{t}}/dM_{t\bar{t}}$ [in fb/GeV].    Pink lines: the SM  with  $Z^\prime$ exchange  prediction for  ($a=b= 0.245$, and $c=d = 0.148$). The same conventions as in  Fig.~\ref{fig.AFBonlya}  used for other lines. }
 \label{fig.RHAFBaeqpbceqd}
\end{figure}

Finally, we consider the case where all couplings are of the same order.  We choose the representative values of the couplings $ a=-b= |g^{L}_{tu}|= 0.239$, and $ c = d = 0.148$. Again
this value for $g^L_{tu}$ satisfies the $|\Delta_d|$ constraint but not the phase $\phi^{\Delta}_d$ constraints from $B_d$ mixing (see Fig.~\ref{fig.Bdconst}).
In Fig.~\ref{fig.AFBaeqmbceqd}, we show the $M_{t\bar{t}} $ distribution for the $t\bar{t}$ observables $A^{t\bar{t}}_{FB}$, and $\sigma_{t\bar{t}}$. We note that
 $A^{t\bar{t}}_{FB}$ can be explained within one $\sigma$ error of its measurement for $M_{t\bar{t}} > 450 $ GeV. The distribution $d\sigma_{t\bar{t}}/dM_{t\bar{t}}$  is also consistent with the measurements. Similar results are obtained with $  a=b= |g^{R}_{tu}|= 0.245$, and $c = d= 0.148$ as shown in Fig. \ref{fig.RHAFBaeqpbceqd}.  The conclusion is that the inclusion of the tensor couplings does not have a significant effect on the top $\AFB$ and can only slightly lower the values of the couplings $a$ and $b$ relative to their values in the pure case, with no tensor couplings, discussed earlier. The presence of the tensor couplings may have an important impact on the polarization measurement in $t \bar{t}$ production \cite{debo}.   

\section{ $t \to u Z^\prime$ Branching ratio}
In this section we consider the decay width for $t \rightarrow u Z'$. 
The decay width with the most general $tuZ'$ coupling is given as, 
\bea  
\Gamma (t\rightarrow u Z^\prime)  & = &
\frac{1}{16       \pi       m_t}      \left(1-{m_{Z^\prime}^2\over       m_t^2}
\right)\left({m_t^2\over             m_{Z^\prime}^2}-1\right)        \Big[
(m_t^2+2m_{Z^\prime}^2)(a^2+b^2) \nl  & &  - 6m_{Z^\prime}^2
( a c-b d)   + m_{Z^\prime}^2({m_{Z^\prime}^2\over m_t^2}+2)(c^2+d^2)  \Big]. \
\eea
Branching ratio is defined as
\bea
\label{BR}
 BR_{tuZ^\prime}  & =  &  \frac{\Gamma[t \to  c Z^\prime]}{\Gamma[m_t]}.
 \eea
 For  the top  width we   use $
\Gamma(m_t) \approx \Gamma (t\rightarrow bW) $ which is given by, \bea
\Gamma  (t\rightarrow bW)  &  = &  {G_F\over 8\pi\sqrt{2}}  |V_{tb}|^2
m_t^3     \left(1-{m_W^2\over     m_t^2}\right)    \left(1+{m_W^2\over
m_t^2}-2{m_W^4\over m_t^4}\right) .\
\label{top_width}
\eea

\begin{figure}[htb!]
\centering
\includegraphics[width=5.5cm]{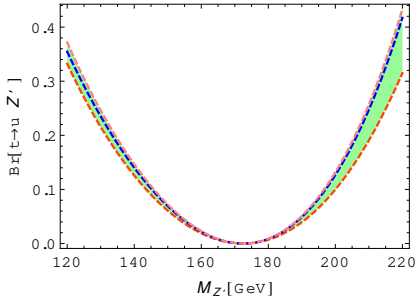} 
\caption{$BR_{Z^\prime} $ vs $M_{Z^\prime}$.  Red dashed line is for   $a = \pm b = 0.257$, and c = d =0. Blue dashed line is for a = -b = 0.239, c = -d = 0.148. Pink dashed line is for a = b = 0.246, c = d = 0.148. }
 \label{fig.BrtuZp}
\end{figure}

In Fig.~\ref{fig.BrtuZp} we show the variation of $t \to u Z^\prime$  branching ratio with $M_{Z^\prime}$ for different couplings.  For couplings $a = \pm b = 0.257$, and $c = d =0$  (red dashed line), we get $BR_{tuZ^\prime} \sim 6\%$ at $m_t = 172.5$ GeV, for a = - b = 0.239, c = -d = 0.148 (blue dashed line),  $BR_{tuZ^\prime}$ is $ 6.9\%$, and  for  a = b = 0.246, c = d = 0.148 (pink dashed line), $BR_{tuZ^\prime}$ is $7.2\%$. These branching ratios may be observable at the LHC \cite{zurek}.

\section{Conclusion}
A large forward-backward asymmetry  in $\ttbar$ production, about a $3.4 \sigma$ away from the SM prediction, has been reported by the CDF collaboration. A  $Z^\prime$ with flavor-changing $tuZ^\prime$ coupling can explain this anomaly. In this work we considered $B_{d,s}$ constraints on the $t q^\prime Z^\prime$ couplings ( $q^\prime=u,
t$). These constraints resulted from the bounds on the effective $b(s,d) Z'$ vertices generated from vertex corrections involving the $tuZ^\prime$ couplings. We found that the right-handed couplings were generally not  tightly constrained but the left-handed couplings were tightly bound from the $B_{d,s}$ mixing data. We then considered the most general $tuZ^\prime$ coupling including tensor terms and found that the tensor terms did not affect the top $\AFB$ in a significant manner. Finally we computed the branching ration for the $t \to u Z^\prime$ transition and found it to be in the percentage range.

\bigskip
\noindent
{\bf Acknowledgments}: This work was financially
supported by the US-Egypt Joint Board on Scientific and Technological
Co-operation award (Project ID: 1855) administered by the US
Department of Agriculture.

\appendix
\section{Functions in scattering amplitudes}
\label{sampfun}
For the scattering amplitudes calculation in $t\bar{t}$ center-of-mass frame, we choose the  relevant coordinates of particle momenta as 
\bea
\label{mom-par}
p_{q,\bar{q}} &=& \frac{\hat{s}}{2}\, (1, 0, 0, \pm 1)\,, \nl
p_{t,\bar{t}} &=& \frac{\hat{s}}{2}\, (1, \pm \beta_t \sin{\theta}, 0, \pm \beta_t \cos{\theta}).
\eea.

With this choice and assume all the couplings in Eq.~(\ref{Lagtuzp-def}) to be real, we obtain the functions $f_i$ in the scattering amplitude in Eq.~(\ref{Sc-amp})  as

\bea
\label{sampfun-int}
f_1 &=& \frac{\hat{s}}{2} \Big[8 \Big(2 a^2 + 2 b^2 + a c - c^2 + 3 b d + d^2 \Big) \frac{m_t^2}{\hat{s}}  + 
   2 \Big(2 a^2 (1 + c_\theta)^2   + 2 b^2 (1 + c_\theta)^2  \nl && + 
      b d (-7 + 4  c_\theta + 6  c_\theta^2 - 3 \beta_t^2) - (c^2 - 
         d^2) (-1 + 3  c_\theta^2 - 2 \beta_t^2) + 
      a c (-1 + \beta_t^2) \Big) 
 \nl && - \Big((-1 +  c_\theta) (c^2 - 
      d^2) (-1 + 2  c_\theta +  c_\theta^2 - 2 \beta_t^2) 
\hat{s}^2 \Big) \frac{\hat{s} }{m_t^2} \Big]\, , \nl
f_2 &=& - \Big(\frac{m_t^2}{\hat{t}}\Big) \hat{s} (a^2 + b^2) 
\Big[(-1 +  c_\theta)^2 + \frac{4 m_t^2}{\hat{s}} \Big].
\eea

\bea
f_3 &=& \frac{1}{16} \hat{s}^2 \Big[32 (a^4 + b^4) \Big(3 + 2 c_\theta + c_\theta^2 + 
       2 \beta_t^2 \Big) +   \frac{1}{m_t^4} (c^4 + d^4)  \Big(32 (9 - 2  c_\theta +  c_\theta^2) m_t^4 \nl && + 
   32 (-5 + 3 c_\theta +  c_\theta^2 +  c_\theta^3) m_t^2 \hat{s} + 
  \hat{s}^2 (5 - 2  c_\theta^2 + \beta_t^2 - 
      c_\theta (3 + \beta_t^2))^2 \Big) + 128 a^3 c \Big(-2 c_\theta  + c_\theta ^2 + \beta_t^2\Big)\nl && -\frac{1}{m_t^4}
 2 c^2 d^2 \Big(-32 (-5 + 3 c_\theta + c_\theta^2 + 
       c_\theta^3) m_t^2  \hat{s} + 
    32 m_t^4 (-11 + 6 c_\theta + c_\theta^2 - 4 \beta_t^2)\nl && - 
     \hat{s}^2 (5 - 2 c_\theta^2 + \beta_t^2 - 
      c_\theta (3 + \beta_t^2))^2 \Big) + \frac{16}{m_t^2} a c \Big(8 b^2 m_t^2 (-2 + 2 c_\theta + 3 c_\theta^2 - 
      3 \beta_t^2)\nl && + 
   c^2 (-1 +c_\theta) (8 (-3 + c_\theta) m_t^2 + 
      \hat{s} (5 + 2 c_\theta^2 - 3 \beta_t^2 - 
        c_\theta (3 + \beta_t^2))) - 
   d^2 (8 m_t^2 (-5 + 8 c_\theta + c_\theta^2 - 4 \beta_t^2) \nl && - (-1 +
         c_\theta) \hat{s} (5 + 2 c_\theta^2 - 3 \beta_t^2 - 
        c_\theta (3 + \beta_t^2)))\Big) + \frac{16}{m_t^2} b^2 \Big(c^2 (2 m_t^2 (-11 - 2  c_\theta + 5  c_\theta^2 - 
         8 \beta_t^2)\nl && - (-1 +  c_\theta) \hat{s} (5 + 3 \beta_t^2 + 
          c_\theta (7 + \beta_t^2))) - 
   d^2 (4 m_t^2 (7 - 2  c_\theta + 2  c_\theta^2 + \beta_t^2) \nl &&+ (-1 + 
          c_\theta) \hat{s} (5 + 3 \beta_t^2 + 
         c_\theta (7 + \beta_t^2)))\Big) + \frac{16}{m_t^2} a^2 \Big(4 b^2 m_t^2 (1 + 6 c_\theta + 3 c_\theta^2 -   2 \beta_t^2)\nl && - 
   d^2 (2 m_t^2 (3 + 18 c_\theta + 3 c_\theta^2 - 
         8 \beta_t^2) + (-1 + c_\theta) \hat{s} (5 + 3 \beta_t^2 + 
         c_\theta (7 +\beta_t^2))) \nl &&+ 
   c^2 (5 \hat{s} + 4 m_t^2 \beta_t^2 + 3 \hat{s} \beta_t^2 - 
      2 c_\theta (24 m_t^2 + \hat{s} (-1 + \beta_t^2)) + 
     c_\theta^2 (12 m_t^2 - \hat{s} (7 + \beta_t^2)))\Big)\Big]\,,\nl
f_4 &=& -\frac{1}{2}\Big(\frac{m_t^2}{\hat{t}}\Big) \hat{s} \Big[32 (a^4 + b^4 + 2 a^3 c + 2 a b^2 c + a^2 (2 b^2 + c^2) - 
      b^2 d^2) m_t^2  + 
   8 (-3 + 2 c_\theta + c_\theta^2)\nl && (a^3 c + a b^2 c + a^2 c^2 - 
      b^2 d^2) \hat{s} + \frac{1}{m_t^2} (-1 + c_\theta)^2 (a^2 c^2 - b^2 d^2) \hat{s}^2 (5 + 
        2 c_\theta + \beta_t^2) \Big]\,,\nl
f_5 &=& \Big(\frac{m_t^2}{\hat{t}}\Big)^2  \hat{s}^2  (a^2 + b^2)^2 (-1 + c_\theta)^2.
 \eea


\end{document}